\documentclass{cpbtex}
\usepackage{graphicx}
\usepackage{bm}

\begin{document}
\begin{CJK*}{GBK}{song}

\title{Faraday rotations, ellipticity and circular dichroism in the magneto-optical spectrum of moir\'{e} superlattices\thanks{Project supported by the National Natural Science Foundation of China (Grant No.~\textcolor{red}{xxxxxxxx}) and Science and
Technology Commission of Shanghai Municipality (Grant No.~19ZR1436400)}}

\author{J. A. Crosse$^{1,2}$ \ and \ Pilkyung Moon$^{1,2,3}$\thanks{Corresponding author. E-mail:~pilkyung.moon@nyu.edu}\\
$^{1}${Arts and Sciences, New York University Shanghai, Shanghai, 200122, China.}\\
$^{2}${NYU-ECNU Institute of Physics at NYU Shanghai, Shanghai, 200062, China.}\\
$^{3}${Department of Physics, New York University, New York, 10003, USA.}}

\date{\today}
\maketitle

\begin{abstract}
We study the magneto-optical conductivity of a number of Van der Waals heterostructures, namely, twisted bilayer graphene, AB-AB and AB-BA stacked twisted double bilayer graphene and monolayer graphene and AB-stacked bilayer graphene on hexagonal boron nitride.
As magnetic field increases, the absorption spectrum exhibits a self-similar
recursive pattern reflecting the fractal nature of the energy spectrum.
Whilst twisted bilayer graphene displays only weak circular dichroism,
monolayer graphene and AB-stacked bilayer graphene on hexagonal boron nitride show specifically strong circular dichroism, owing to strong inversion symmetry breaking properties of the hexagonal boron nitride layer.
As, the left and right circularly polarized light interact with these structures differently, plane polarized incident light undergoes a Faraday rotation and gains an ellipticity when transmitted. The size of the respective angles is on the order of a degree.
\end{abstract}



\section{Introduction}

Van der Waals heterostructures \cite{vdw1, vdw2, vdw3, vdw4, vdw5}, stacks of $2D$ monolayers with relative twist angles between them, offer a novel way to engineer materials with specific electronic properties and, hence, are a promising platform for high-performance electronics. As the optical properties of a material are directly related to its electronic properties, these types of heterostructure potentially offer a method for optical control as well. One immediate concern would be optical depth. However, although only a few atoms in thickness, the quantum efficiency of these structures can be remarkably high - monolayer graphene displays a universal absorption of $2.3\%$ \cite{gabs} and recent experiments in hexagonal Boron Nitride encapsulated graphene have shown $36\%$ absorption for mid-infrared frequencies when a perpendicular magnetic field is applied \cite{ghbnabs}. Furthermore, single and multilayer graphene has also been shown to display giant Faraday rotations of up to $6\,\mathrm{deg.}$ when subjected to a magnetic field \cite{farad4}, another feature that could potentially be exploited for optical control. These results lead one to conjecture as to whether van der Waals heterostructures subject to an applied magnetic field might be suitable material for optical device design.

\begin{figure}[t]
\centering
\includegraphics[width=0.65\columnwidth]{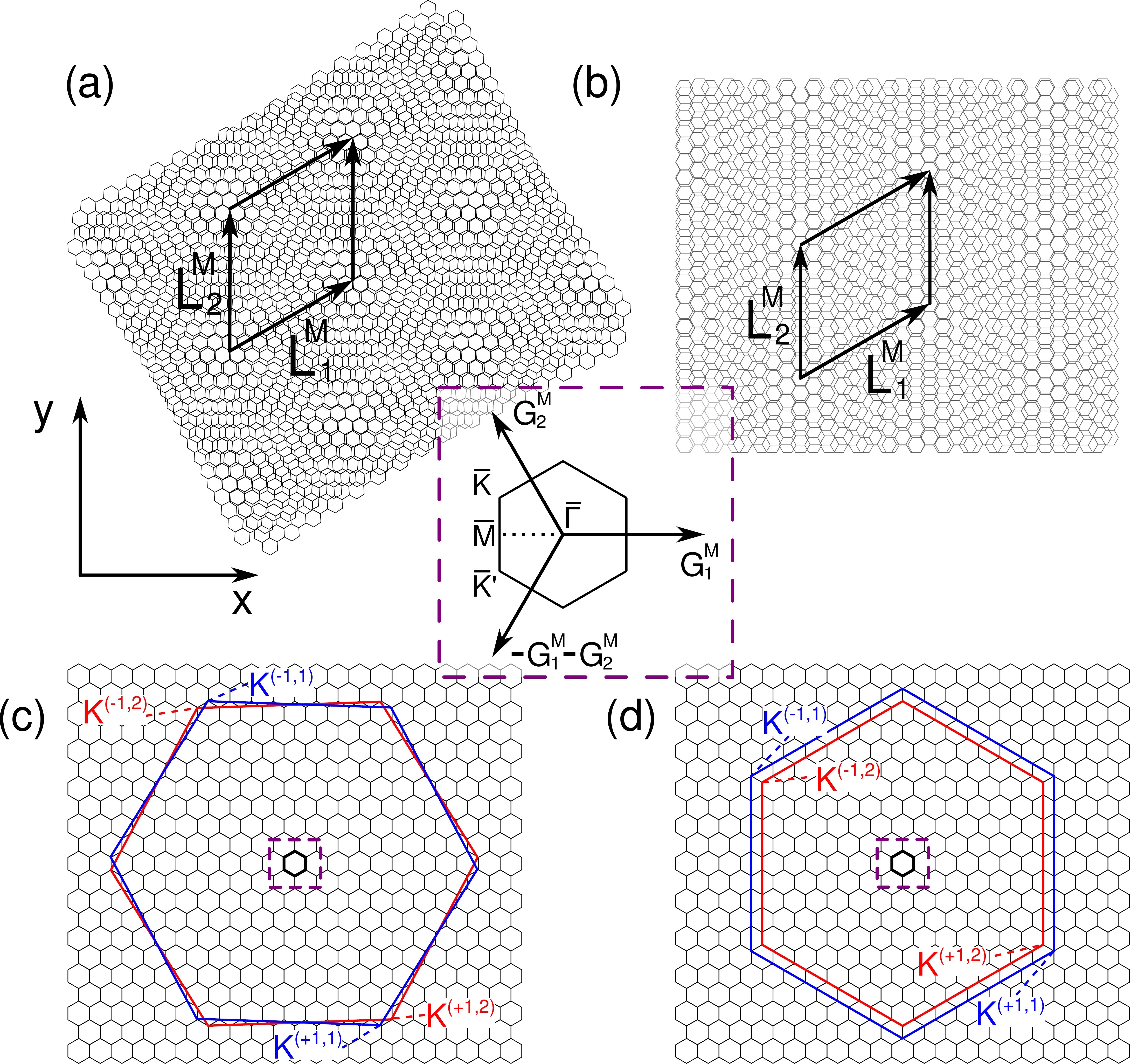}
\caption{The moir\'{e} superlattice generated by (a) two layers with different orientations and (b) two layers with different lattice constants. Brillouin zone folding in moir\'{e} superlattice generated by (c) two layers with different orientations and (d) two layers with different lattice constants. The large red and blue hexagons represent the Brillouin zone of the two layers. The small hexagons represent the moir\'{e} Brillouin zone. The inset shows the high-symmetry points of the moir\'{e} Brillouin zone.}
\label{fig1}
\end{figure}

In addition to potential technological application, van der Waals heterostructures also provide a direct probe of fundamental physics. When a material is subjected to an applied magnetic field the electron band structure splits up into a series of discrete Landau levels. At higher magnetic fields, specifically when the magnetic length becomes smaller than the lattice constant, the individual Landau levels split into a series of sublevels resulting in the fractal-like Hofstadter spectrum \cite{hof}. In traditional materials, the magnetic field required to realize the Hofstadter spectrum is unfeasibly large - on the order of $10^5\,\mathrm{T}$. However, in van der Waals heterostructures the lattice constant of the moir\'{e} interference pattern, which is generated when adjacent layers have different orientations and/or different lattice constant (see Fig.~\ref{fig1}), can be very large and hence the Hofstadter regime can be realized at reasonable magnetic fields \cite{tbghof, ghbnhof}. This led to it's indirect observation via measurements of the Hall conductivity \cite{hofobs1, hofobs2, hofobs3, hofobs4}. The optical conductivity offers a way to directly observe the energy spacing between bands and, hence, could be used to observe the fractal nature of the Hofstadter spectrum directly \cite{tbgmo}.

Here, we present the magneto-optical spectrum, computed via the optical conductivity, of a variety of van der Waals Heterostructures. The specific structures we consider are: twisted bilayer graphene (TBG), AB-AB and AB-BA stacked twisted double bilayer graphene (TDBG) and monolayer graphene and AB-stacked bilayer graphene on hexagonal boron nitride (MLG-hBN and BLG-hBN respectively). In all but the first structure the materials display strong circular dichroism, where the optical conductivity is significantly different for the two different polarizations of circularly polarized light. Some materials also display valley dichroism, where the contribution to the optical conductivity from each valley is significantly different. This would allow optical excitation of a single valley and, hence, may provide a suitable platform for `valleytronic' devices \cite{valley1}. We also compute the circular dichroism induced Faraday rotations and ellipticities experienced by transmitted plane-polarized light. At high magnetic fields the Hofstadter spectrum is clearly visible in the optical conductivity and hence would correspond to a direct probe of the fractal nature of bands. 

\begin{figure}[t]
\centering
\includegraphics[width=0.5\columnwidth]{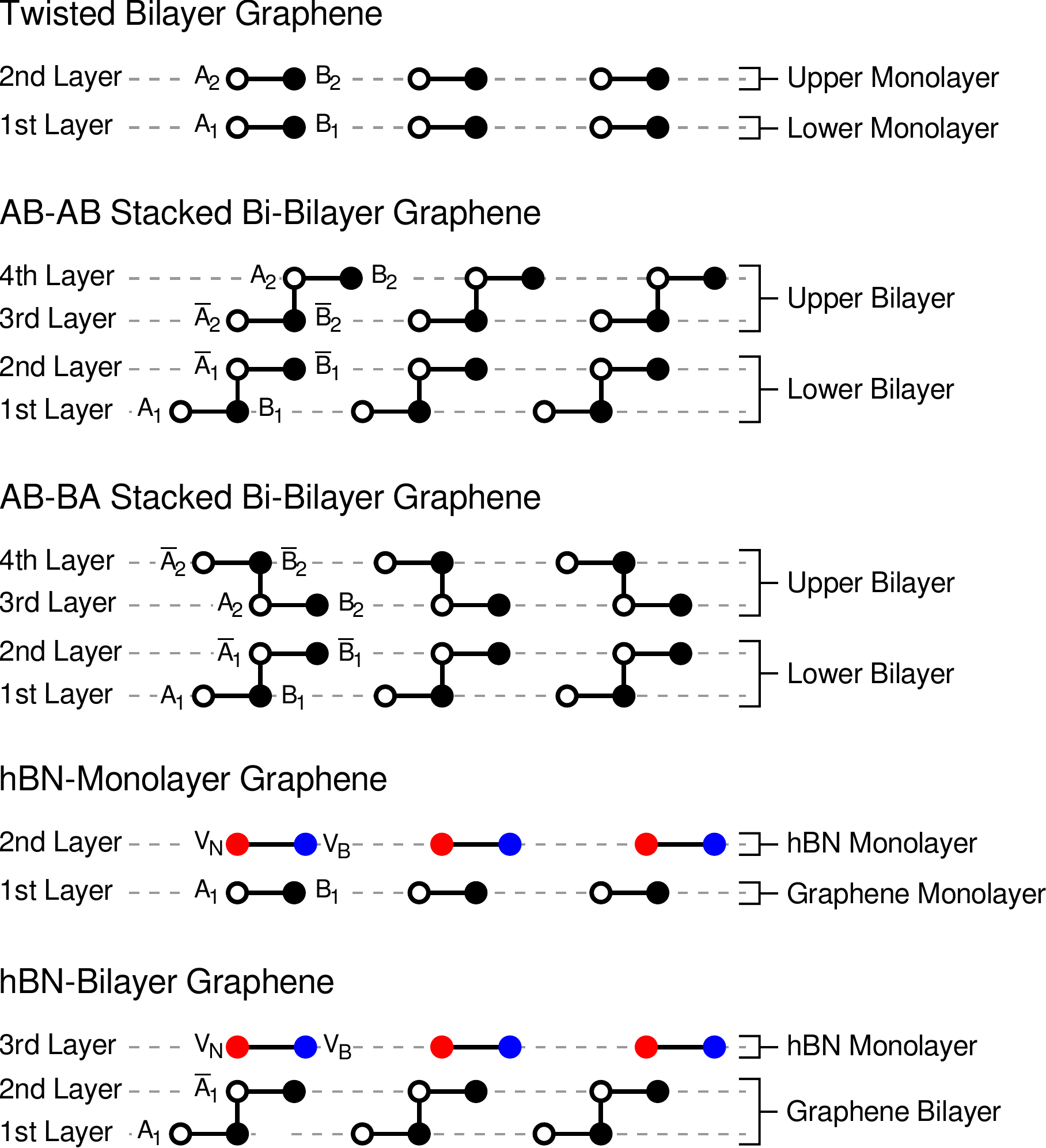}
\caption{Cross-section through the various van der Waals heterostructure considered here.}
\label{fig2}
\end{figure}

\section{Theoretical Methods}

In order to study the optical conductivity of van der Waals heterostructures we will use an effective continuum model. This model is valid when the moir\'{e} lattice constant is much larger than the graphene lattice constant \cite{em1, em2, em3, em4, em5, em6, em7, em8}. In such systems, the separation between the Dirac points is sufficiently large that inter-valley mixing can be safely neglected and the full Hamiltonian separates into two independent Hamiltonians, each describing the electronic properties of a single valley.

\subsection{Atomic Structure of Monolayer Graphene}

Monolayer graphene (MLG) consists of a hexagonal lattice of carbon atoms, each contributing a single p-orbital electron to the electronic structure. In the following, we take the lattice vectors of the unrotated graphene layers to be $\mathbf{a}_{1} = a(1,0)$ and $\mathbf{a}_{2} = a(1/2,\sqrt{3}/2)$ with the graphene lattice constant $a=0.246\,\mathrm{nm}$. In twisted structures the two layers are rotated by an angle $\theta$ with respect to each other. After rotation, the lattice vectors in each layer read $\mathbf{a}_{i}^{(l)} = R(\mp\theta/2) \mathbf{a}_{i}$ where $R(\theta)$ is the rotation matrix and $l \in 1, 2$ refers to rotations of $\mp\theta$, respectively. Accordingly, the unrotated reciprocal lattice vectors are given by $\mathbf{a}_{1}^{\ast} = (2\pi/a)(1,-1/\sqrt{3})$ and $\mathbf{a}_{2}^{\ast} = (2\pi/a)(0,2/\sqrt{3})$ and the rotated reciprocal lattice vectors in each layer by $\mathbf{a}_{i}^{\ast(l)} = R(\mp\theta/2) \mathbf{a}_{i}^{\ast}$. The Dirac points of the two graphene layers are located at $\mathbf{K}^{(\xi,l)}=-\xi[2\mathbf{a}_{1}^{\ast(l)}+\mathbf{a}_{2}^{\ast(l)}]/3$, with $\xi=\pm 1$ labeling the $K$ and $K'$ valleys. The Hamiltonian for each valley reads
\begin{equation}
H_{MLG}^{(\xi,l)}(\mathbf{k}^{(\xi,l)}) = \left(\begin{array}{cc}
0 & -\hbar v_{F}k^{(\xi,l)}_{-}\\
-\hbar v_{F}k^{(\xi,l)}_{+} & 0
\end{array}\right).
\label{H_MLG}
\end{equation}
Here, $v_{F}$ is a Fermi velocity, which we take to be $\approx 1\times10^6\,\mathrm{m/s}$ in MLG, $k^{(\xi,l)}_{\pm} = e^{\pm i\xi\eta^{(l)}}(\xi k_{x}^{(\xi,l)}\pm ik_{y}^{(\xi,l)})$, where $\eta^{(l)}$ is the angle between $\mathbf{a}_{1}^{(l)}$ and the $x$-axis (which in this case is $\eta^{(1)/(2)} = \pm \theta/2$), with the momentum taken in relation to the Dirac points [$\textbf{k}^{(\xi,l)} = \textbf{k}-\textbf{K}^{(\xi,l)}$].

\subsection{Atomic Structure of Bilayer Graphene}

In the following we will consider two possible stackings of bilayer graphene (BLG), AB- and BA- stacking. In isolation, these two stackings are identical - the two forms being related by a 180 degree in-plane rotation. However, when part of a van der Waals heterostructure these two stackings are distinguishable because the atoms in the BLG layer are in different locations with respect to the atoms of the adjoining layers. The BLG Hamiltonian for each valley reads
\begin{equation}
H_{AB}^{(\xi,l)}(\mathbf{k}^{(\xi,l)}) = \left(\begin{array}{cc}
H_{MLG}^{(\xi,l)}(\mathbf{k}^{(\xi,l)})+\Delta_{B} & g^{\dagger}(\mathbf{k}^{(\xi,l)})\\
g(\mathbf{k}^{(\xi,l)}) & H_{MLG}^{(\xi,l)}(\mathbf{k}^{(\xi,l)})+\Delta_{A}\\
\end{array}\right),
\label{H_AB}
\end{equation}
for AB-stacking and
\begin{equation}
H_{BA}^{(\xi,l)}(\mathbf{k}^{(\xi,l)}) = \left(\begin{array}{cc}
H_{MLG}^{(\xi,l)}(\mathbf{k}^{(\xi,l)})+\Delta_{A} & g(\mathbf{k}^{(\xi,l)})\\
g^{\dagger}(\mathbf{k}^{(\xi,l)}) & H_{MLG}^{(\xi,l)}(\mathbf{k}^{(\xi,l)})+\Delta_{B}\\
\end{array}\right),
\label{H_BA}
\end{equation}
for BA-stacking. Here, $H_{MLG}^{(\xi,l)}(\mathbf{k}^{(\xi,l)})$ is the MLG Hamiltonian given in Eq. \eqref{H_MLG},
\begin{equation}
g(\mathbf{k}_l) = \left(\begin{array}{cc}
\hbar v_{4}k^{(\xi,l)}_{+} & \gamma_{1}\\
\hbar v_{3}k^{(\xi,l)}_{-} & \hbar v_{4}k^{(\xi,l)}_{+}
\end{array}\right),
\label{BLGint}
\end{equation}
is the intra-bilayer coupling and
\begin{equation}
\Delta_{A} = \left(\begin{array}{cc}
\Delta & 0\\
0 & 0
\end{array}\right), \qquad \Delta_{B} = \left(\begin{array}{cc}
0 & 0\\
0 & \Delta
\end{array}\right),
\end{equation}
are the on-site potential of the lattice sites that are vertically aligned with the lattice sites of the adjacent layer of the bilayer. Numerically, we take this to be $\Delta = 0.05\,\mathrm{eV}$. $v_{F}$ is, again, the Fermi velocity which, we take to be $\approx 0.8\times10^6\,\mathrm{m/s}$ in BLG to make the results consistent with previous work \cite{tdbg, tdbghof}. The remaining parameters are $\gamma_{1} = 0.4\,\mathrm{eV}$, the interaction strength between the $A$ and $B$ sites in the upper and lower layers of the graphene bilayer and $v_3 = 1.036\times 10^5\,\mathrm{m/s}$ and $v_4 = 0.143\times 10^4\,\mathrm{m/s}$, which are responsible for the trigonal warping and electron-hole asymmetry, respectively \cite{tdbg, tdbghof, tdbgparam}.

\subsection{Interlayer Coupling Between Graphene Layers}

When two $2D$ materials with different lattice constants and/or different orientations are stacked on top of each other, the mismatch between the lattices of the two layers leads to a moir\'{e} interference pattern. In the low energy regime, the long wavelength of the electrons are more strongly affected by the long period moir\'{e} superlattice rather than the short period of the individual lattices of each layer. The dominance of the moir\'{e} superlattice leads to a dramatic increase in the size of the material's unit cell and a correspondingly large reduction in the size of the Brillouin zone. The reciprocal lattice vectors for the moir\'{e} Brillouin zone are given by $\mathbf{G}_{i}^{\mathrm{M}}=\mathbf{a}_{i}^{\ast(1)}-\mathbf{a}_{i}^{\ast(2)}$, from which the real space moir\'{e} lattice vectors can be found via $\mathbf{G}_{i}^{\mathrm{M}}\cdot\mathbf{L}_{j}^{\mathrm{M}} = 2\pi$. The moir\'{e} lattice constant is given by $L_{\mathrm{M}} = a/[2\sin(\theta/2)]$ and the area of the moir\'{e} unit cell by $A=|\mathbf{L}_{1}^{\mathrm{M}}\times\mathbf{L}_{2}^{\mathrm{M}}| = (\sqrt{3}/2)L_{\mathrm{M}}^{2}$. The interlayer interaction can be found by considering the points in the Brillouin zone coupled by the generalized Umklapp process \cite{em6}. By considering only the three strongest processes (the next strongest processes have coupling strengths two orders of magnitude smaller), the form of the coupling between adjacent graphene layers can be found to be
\begin{equation}
U = \left(\begin{array}{cc}
u & u'\\
u' & u
\end{array}\right) +
\left(\begin{array}{cc}
u & u'\omega^{-\xi}\\
u'\omega^{\xi} & u
\end{array}\right)e^{i\xi\mathbf{G}_{1}^{\mathrm{M}}\cdot\mathbf{r}}
+ \left(\begin{array}{cc}
u & u'\omega^{\xi}\\
u'\omega^{-\xi} & u
\end{array}\right)e^{i\xi(\mathbf{G}_{1}^{\mathrm{M}}+\mathbf{G}_{2}^{\mathrm{M}})\cdot\mathbf{r}},
\label{U}
\end{equation}
where $\omega = e^{2\pi i/3}$ and $u = 0.07974\,\mathrm{eV}$ and $u' = 0.09754\,\mathrm{eV}$ are coupling constants that give the strength of the interaction between like ($A \leftrightarrow A$, $B \leftrightarrow B$) and opposing ($A \leftrightarrow B$) sublattices, respectively \cite{em8}. Thus, one can write down the general form of the Hamiltonian for TBG
\begin{equation}
H_{TBG}^{(\xi)} = \left(\begin{array}{cc}
H_{MLG}^{(\xi,1)}(\mathbf{k}^{(\xi,1)}) & U^{\dagger}\\
U & H_{MLG}^{(\xi,2)}(\mathbf{k}^{(\xi,2)})\\
\end{array}\right),
\end{equation}
and, similarly, for TDBG
\begin{equation}
H_{ij}^{(\xi)} = \left(\begin{array}{cc}
H_{i}^{(\xi,1)}(\mathbf{k}^{(\xi,1)}) & U_{4\times 4}^{\dagger}\\
U_{4\times 4} & H_{j}^{(\xi,2)}(\mathbf{k}^{(\xi,2)})\\
\end{array}\right),
\end{equation}
where $i, j \in AB, BA$ and
\begin{equation}
U_{4\times 4} = \left(\begin{array}{cc}
0 & U\\
0 & 0
\end{array}\right).
\end{equation}

\subsection{The Potential Produced by Hexagonal Boron Nitride}

Hexagonal Boron Nitride (hBN) is another $2D$ material that has a hexagonal lattice. However, in this material the two triangular sublattices consist of different atomic species. As a result, inversion symmetry is broken and hBN exhibits a $4.7\,\mathrm{eV}$ gap. In a van der Waals heterostructures where graphene and hBN layers are adjacent, the low energy band structure is dominated by the graphene bands and the contribution of the hBN layer can be treated as a perturbing potential, the form of which can be identified by eliminating the hBN basis using second order perturbation theory \cite{ghbnhof}. By neglecting the dispersion in the hBN Hamiltonian, which is valid for small twist angle, $\theta$, when the K-points in the adjacent layers are close to each other, one finds that  
\begin{equation}
H_{hBN} = \left(\begin{array}{cc}
V_{N} & 0\\
0 & V_{B}
\end{array}\right),
\label{hBN}
\end{equation}
with $V_{N} = 3.34\,\mathrm{eV}$ and $V_{B} = -1.40\,\mathrm{eV}$. Using this simplified Hamiltonian in Eq. \eqref{hBN} and the parameters $u_{0} = u = u' = 0.152\,\mathrm{eV}$ in Eq. \eqref{U}, one can identify the potential at $E\approx 0$ as
\begin{align}
V_{hBN} &= U^{\dagger}\left(-H_{hBN}\right)U,\nonumber\\
&= V_{0}\left(\begin{array}{cc}
1 & 0\\
0 & 1
\end{array}\right) + \left\{V_{1}e^{i\xi\psi}\left[\left(\begin{array}{cc}
1 & \omega^{-\xi}\\
1 & \omega^{-\xi}
\end{array}\right)e^{i\xi\mathbf{G}_{1}^{\mathrm{M}}\cdot\mathbf{r}}\right.\right.\nonumber\\
&\qquad\left.\left.  +
\left(\begin{array}{cc}
1 & \omega^{\xi}\\
\omega^{\xi} & \omega^{-\xi}
\end{array}\right)e^{i\xi\mathbf{G}_{2}^{\mathrm{M}}\cdot\mathbf{r}} + \left(\begin{array}{cc}
1 & 1\\
\omega^{-\xi} & \omega^{-\xi}
\end{array}\right)e^{-i\xi(\mathbf{G}_{1}^{\mathrm{M}}+\mathbf{G}_{2}^{\mathrm{M}})\cdot\mathbf{r}}\right] + \mathrm{h.c.}\right\},
\label{VhBN}
\end{align}
with
\begin{gather}
V_{0} = -3u_{0}^{2}\left(\frac{1}{V_{N}}+\frac{1}{V_{B}}\right),\\
V_{1}e^{i\psi} = -u_{0}^{2}\left(\frac{1}{V_{N}}+\omega\frac{1}{V_{B}}\right).
\end{gather}
The numerical values of the parameters are given by $V_{0}\approx 0.0289\,\mathrm{eV}$, $V_{1}\approx 0.0210\,\mathrm{eV}$ and $\psi \approx -0.29~\mathrm{(rad)}$ \cite{ghbnhof}. Thus, the Hamiltonians for MLG-hBN and BLG-hBN heterostructures are
\begin{equation}
H_{MLG-hBN}^{(\xi)} = H_{MLG}^{(\xi)}(\mathbf{k}) + V_{hBN},
\end{equation}
and
\begin{equation}
H_{BLG-hBN}^{(\xi)} = H_{AB}^{(\xi)}(\mathbf{k}) + V_{hBN, 4\times 4}
\end{equation}
respectively, with
\begin{equation}
V_{hBN, 4\times 4} = \left(\begin{array}{cc}
V_{hBN} & 0\\
0 & 0
\end{array}\right).
\end{equation}

\subsection{Van der Waals Heterostructures in Magnetic Fields}

In the absence of a magnetic field, the band structure of the material can be found by diagonalizing the Hamiltonain in the basis of the Bloch wavefunctions for each layer. To find the band structure when a perpendicular magnetic field is applied, the Hamiltonain can be diagonalized in the basis of single particle monolayer graphene Landau levels (in the following the Zeeman effect is neglected) \cite{tbghof, ghbnhof, tdbghof}. In general, in the presence of a magnetic field, the periodicity of the lattice is lost owing to the spatial dependence of the vector potential, which, in the Landau gauge, reads $\mathbf{A} = (0,Bx,0)$. However, for certain values of the magnetic field - specifically, when the number of quanta of magnetic flux per unit cell is a rational number (i.e. $\Phi/\Phi_{0} = p/q$ where $p$ and $q$ are co-prime integers, $\Phi = BA$ is the flux through the unit cell and $\Phi_{0} = h/e$ is the quantum of magnetic flux) - in the Landau gauge with the y-axis of the coordinate system parallel to $\mathbf{L}_{2}$, one can introduce a periodic magnetic unit cell with lattice vectors $\tilde{\mathbf{L}}_{1} = q\mathbf{L}_{1}$ and $\tilde{\mathbf{L}}_{2} = \mathbf{L}_{2}$ \cite{mbzc1, mbzc2}. Hence, one can find `magnetic' Bloch conditions for this enlarged unit cell
\begin{gather}
\Psi_{\mathbf{k}}(\mathbf{r}+\tilde{\mathbf{L}}_{1}) = e^{i\mathbf{k}\cdot\tilde{\mathbf{L}}_{1}}e^{-i(e/\hbar)(\mathbf{A}-\mathbf{B}\times\mathbf{r})\cdot\tilde{\mathbf{L}}_{1}}\Psi_{\mathbf{k}}(\mathbf{r}),\label{mbc1}\\
\Psi_{\mathbf{k}}(\mathbf{r}+\tilde{\mathbf{L}}_{2}) = e^{i\mathbf{k}\cdot\tilde{\mathbf{L}}_{2}}\Psi_{\mathbf{k}}(\mathbf{r}).\label{mbc2}
\end{gather}
Therefore, it is only at specific values of the magnetic field that the band structure can be found.

Within each layer, one can construct a wavefunction that obeys the magnetic Bloch conditions in Eqs. \eqref{mbc1} and \eqref{mbc2} from the Landau levels of monolayer graphene. The effective continuum Landau levels for the +ve ($\xi=+1$) and -ve ($\xi=-1$) in monolayer graphene read \cite{ando, shonando}
\begin{gather}
\Psi^{(+,l)}_{n,k_{y}}(\mathbf{r}) = C_{n}e^{ik_{y}y}\left(\begin{array}{c}
-i\mathrm{sgn}(n)\phi_{|n|-1,k_{y}}(x)\\
-e^{i\eta^{(l)}}\phi_{|n|,k_{y}}(x)
\end{array}\right)e^{i\mathbf{K}^{(+,l)}\cdot\mathbf{r}},\\
\Psi^{(-,l)}_{n,k_{y}}(\mathbf{r}) = C_{n}e^{ik_{y}y}\left(\begin{array}{c}
e^{i\eta^{(l)}}\phi_{|n|,k_{y}}(x)\\
-i\mathrm{sgn}(n)\phi_{|n|-1,k_{y}}(x)
\end{array}\right)e^{i\mathbf{K}^{(-,l)}\cdot\mathbf{r}},
\end{gather}
respectively, with the upper and lower components of the vector referring to the $A$ and $B$ sublattices respectively. Here, $n$ is the Landau level index, $k_y$ is the wave vector in the $y$-direction and $\eta^{(l)}$ is, again, the angle between $\mathbf{a}_{1}^{(l)}$ and the $x$-axis. The single particle Landau level is defined in terms of the Hermite polynomial, $H_{n}(z)$, as $\phi_{|n|,k_{y}}(x) = (2^{n}n!\sqrt{\pi}l_{B})^{-1/2}e^{-z^2/2}H_{n}(z)$ with $z=(x+k_{y}l_{B}^{2})/l_{B}$ and $l_{B} = \sqrt{\hbar/(eB)}$ \cite{shonando, zhengando}. The normalization coefficient reads $C_{n} = 1$ for $n=0$ and $C_{n} = 1/\sqrt{2}$ for $n\neq 0$. To find a wavefunction that satisfies Eqs. \eqref{mbc1} and \eqref{mbc2} one needs to combine Landau levels at different $k_{y}$ via
\begin{equation}
\Psi^{(\xi,l)}_{n,m} = \sum_{j=-\infty}^{\infty}\alpha^{j}\Psi^{(\xi,l)}_{n,k_{y}^{m}}(\mathbf{r}),
\label{wvfn}
\end{equation}
with
\begin{gather}
\alpha = e^{i(\mathbf{k}-\mathbf{K}^{(\xi,l)})\cdot(\tilde{\mathbf{L}}_{1}-q\tilde{\mathbf{L}}_{2}/2)}e^{i\pi pq(j+1)/2}e^{i\pi qm},\\
k_{y}^{m} = k_{y} - K^{(\xi,l)}_{y}-\frac{2\pi}{L_{y}}(pj+m),
\end{gather}
and the $m$ index running from $0$ to $p-1$.

The electronic properties of any graphene based van der Waals heterostructure in a magnetic field can be found by constructing a matrix Hamiltonian for each valley in the basis of the wavefunctions given in Eq. \eqref{wvfn}
\begin{equation}
H_{n,n',m,m',l,l'}^{(\xi)} = \langle\Psi^{(\xi,l')}_{n',m'}|H^{(\xi)}_{\mu}|\Psi^{(\xi,l)}_{n,m}\rangle,
\label{H_gen}
\end{equation}
with $\mu$ indicating the specific heterostructure. The diagonal elements of the matrix Hamiltonian reduces to 
\begin{equation}
H_{n,n',m,m',l,l}^{(\xi)} = \varepsilon_{n}\delta_{n,n'}\delta_{m,m'},
\end{equation}
where $\varepsilon_{n} = \hbar\omega_{B}\mathrm{sgn}(n)\sqrt{|n|}$ is the single particle Landau level energy with $\omega_{B} = \sqrt{2v_{F}^2eB/\hbar}$ \cite{shonando, zhengando}. The inter-layer matrix elements of $U$ can be evaluated using the identity \cite{me1}
\begin{multline}
\langle\phi_{|n'|,k_{y}'}(x)e^{ik_{y}'y}|e^{i\mathbf{G}\cdot\mathbf{r}}|e^{ik_{y}y}\phi_{|n|,k_{y}}(x)\rangle = \delta_{k_{y}',k_{y}+G_{y}}\sqrt{\frac{\lambda!}{\Lambda!}}\left(\frac{G_{x} + iG_{y}}{|\mathbf{G}|}\right)^{n-n'}\\
\times\left(\frac{i|\mathbf{G}|l_{B}}{\sqrt{2}}\right)^{|n-n'|}e^{-|\mathbf{G}|^{2}l_{b}^{2}/4}e^{-il_{B}^{2}G_{x}(k_{y}' + k_{y})/2}L_{\lambda}^{|n-n'|}\left(\frac{|\mathbf{G}|^2l_{B}^{2}}{2}\right),
\end{multline}
where $\lambda = \mathrm{min}(n,n')$, $\Lambda = \mathrm{max}(n,n')$ and $L^{\alpha}_{\lambda}(x)$ is an associated Laguerre polynomial. This matrix Hamiltonian is unbounded in both $n$ and $j$. However, by applying magnetic Bloch conditions one can see that the state with $j=1$ and $m=0$ is equivalent to the state with $j=0$ and $m=p$ which is just the state $j=0$ and $m=0$ with the addition of a phase. Thus, one only needs to consider a single cycle of $m \in [0, p-1]$ with the appropriate periodic boundary conditions. The $n$ index relates to the energy of the Landau level basis but one can truncate the Hamiltonian at an energy at which the Landau level only weakly affect the low energy spectrum. This cutoff energy must be significantly larger that the interlayer coupling characterized by the coupling constants $u$ and $u'$. This bounded matrix can then be diagonalized to find the electronic structure of the heterostructure in question.

\subsection{The Magneto-Optical Conductivity}

Once the band structure of the material has been obtained, the optical conductivities for left ($\sigma_{+}$) and right ($\sigma_{-}$) circularly polarized light can be found from \cite{tbgmo, optcon1, optcon2}
\begin{equation}
\sigma_{\pm} = \frac{e^2\hbar}{iS}\sum_{\alpha, \beta}\frac{f(\varepsilon_{\alpha}) - f(\varepsilon_{\beta})}{\varepsilon_{\alpha} - \varepsilon_{\beta}}\frac{|\langle\alpha|\hat{v}_{\pm}|\beta\rangle|^2}{\varepsilon_{\alpha} - \varepsilon_{\beta}+\hbar\omega+i\eta},
\end{equation}
where $S$ is the area of the system. Here, $\hat{v}_{\pm} = \hat{v}_{x}\pm\hat{v}_{y}/\sqrt{2}$ with $v_{j} = -(i/\hbar)[\hat{r}_{j}, \hat{H}]$ ($j\in x, y$) is the velocity operator and $\eta = 0.01\,\mathrm{meV}$ is a phenomenological broadening. $\varepsilon_{i}$ and $|i\rangle$ ($i\in \alpha, \beta$) are the eigenenergies and eigenstates which are found from the diagonalization of the Hamiltonian in Eq. \eqref{H_gen}. The function $f(\varepsilon)$ is the Fermi distribution. In this study we take the temperature to be absolute zero, hence $f(\varepsilon) = 1$ for all states below the charge neutral point and $f(\varepsilon) = 0$ for all states above the charge neutral point. Note that the Peierls substitution, which is used to include the the effects of the applied magnetic field, replaces the momentum with the gauge invariant canonical momentum $\hbar\mathbf{k} \rightarrow \hbar\mathbf{k} + e\mathbf{A}$, where $\mathbf{A}$ is the vector potential. However, if we then choose the Landau gauge $\mathbf{A} = (0, Bx, 0)^{T}$ then the velocity operators remain unchanged in the presence of a magnetic field. Furthermore, the interlayer coupling, as it is not a function of the momentum, also does not affect the velocity operators. Thus, the velocity operators are only dependent on the form of the Hamiltonian in each individual layer.

\subsection{Circular Dichroism and it's experimental signatures: Faraday Rotations and Ellipticity}

Circular dichroism is where the optical response of a material to the two polarization of circularly polarized light is different and can be quantified by the parameter
\begin{equation}
D_{c} = \frac{\mathrm{Re}\left[\sigma_{+}\right] - \mathrm{Re}\left[\sigma_{-}\right]}{\mathrm{Re}\left[\sigma_{+}\right] + \mathrm{Re}\left[\sigma_{-}\right]}.
\label{cd}
\end{equation}
This feature can have a significant effect on incident electromagnetic radiation. Furthermore, as the magneto-optical conductivity for circularly polarized light is a complex variable, it has both real and imaginary parts and these parts also affect the propagation of light in different ways. Specifically, for incident plane polarized light, which consists of an equal mixture of the two circularly polarized components, a difference in the real part of the magneto-optical conductivity leads to a change in the relative magnitude of the two components, generating an ellipticity, and a difference in the imaginary part leads to a change in the relative phase of each component, leading to a Faraday rotation of the polarization direction. These effect can be used to probe the circular dichromatic properties of a material.

For an electromagnetic plane wave incident on linearly responding $2D$ electron gas (i.e. the electromagnetic response is of the form $j=\sigma\cdot\mathbf{E}$), Maxwell's equations give the transmission of circularly polarized light as \cite{farad1}
\begin{equation}
t_{\pm} = \frac{2n_{u}}{n_{u}+n_{s}+(\sigma_{\pm}/c\varepsilon_{0})},
\end{equation}
where $n_{s}$ and $n_{u}$ are the refractive indices of the substrate and encapsulating layer respectively (in the following we will consider free standing structures so $n_{s} = n_{u} = 1$). If circular dichroism is present in the material, then the magneto-optical conductivity for the two circular polarizations will be different and, hence, the transmission properties of these two polarizations will also be different. This difference can be probed by using incident plane polarized light, which will undergo a Faraday rotation, $\theta$, and obtain an ellipticity, $\delta$, the magnitudes of which are given by
\begin{gather}
\theta = \frac{1}{2}\left(\mathrm{Arg}\left[t_{+}\right]-\mathrm{Arg}\left[t_{-}\right]\right) \approx \frac{1}{(n_{u}+n_{s})c\varepsilon_{0}}\left(\mathrm{Im}\left[\sigma_{+}\right]-\mathrm{Im}\left[\sigma_{-}\right]\right),\label{theta}\\
\delta = \frac{|t_{+}|-|t_{-}|}{|t_{+}|+|t_{-}|} \approx \frac{1}{(n_{u}+n_{s})c\varepsilon_{0}}\left(\mathrm{Re}\left[\sigma_{+}\right]-\mathrm{Re}\left[\sigma_{-}\right]\right),\label{delta}
\end{gather}
respectively. The approximation on the right hand side valid for the case of $n_{s}+n_{u}>\sigma_{xy}/c\varepsilon_{0}$ and where one has taken terms to first order in the conductivity only \cite{farad2, farad3}. Thus, the ellipticity is a probe of the real part and the Faraday rotation probes the imaginary part of the circular dichroism.

Finally, some materials presented here display Valley dichroism, in that incident light interacts more strongly with one valley than the other. Here, and in a similar manner to the circular dichroism parameter, the Valley dichroism paramter is defined to be
\begin{equation}
D_{v} = \frac{\mathrm{Re}\left[\sigma_{\bar{K}}\right] - \mathrm{Re}\left[\sigma_{\bar{K'}}\right]}{\mathrm{Re}\left[\sigma_{\bar{K}}\right] + \mathrm{Re}\left[\sigma_{\bar{K'}}\right]},
\label{vd}
\end{equation}
where $\sigma_{K(')}$ is the magneto-optical conductivity for the $K(')$ valley only.

\section{Results and Discussion}

\subsection{Twisted Bilayer Graphene}

Under the influence of a perpendicular magnetic field the continuous band structure of MLG splits up into a series of discrete Landau levels with the energy and angular momentum of each level given by $\varepsilon_{n} = \hbar\omega_{B}\mathrm{sgn}\left(n\right)\sqrt{|n|}$ and $L_{n} = \hbar|n|$, where $\omega_{B} = \sqrt{2v_{f}^2eB/\hbar}$ is the cyclotron frequency and $n$ is the Landau level index. Left and right circularly polarized light carries angular momentum of $\pm\hbar$, respectively, and hence can drive transitions between certain Landau levels, specifically those related by $|n| \rightarrow |n|+1$ for left circularly polarized light and $|n| \rightarrow |n|-1$ for right circularly polarized light. These selection rules lead to a discrete set of peaks in the magneto-optical conductivity which correspond to allowed transitions between between specific states.
\begin{figure}[h]
\centering
\includegraphics[width=0.9\columnwidth]{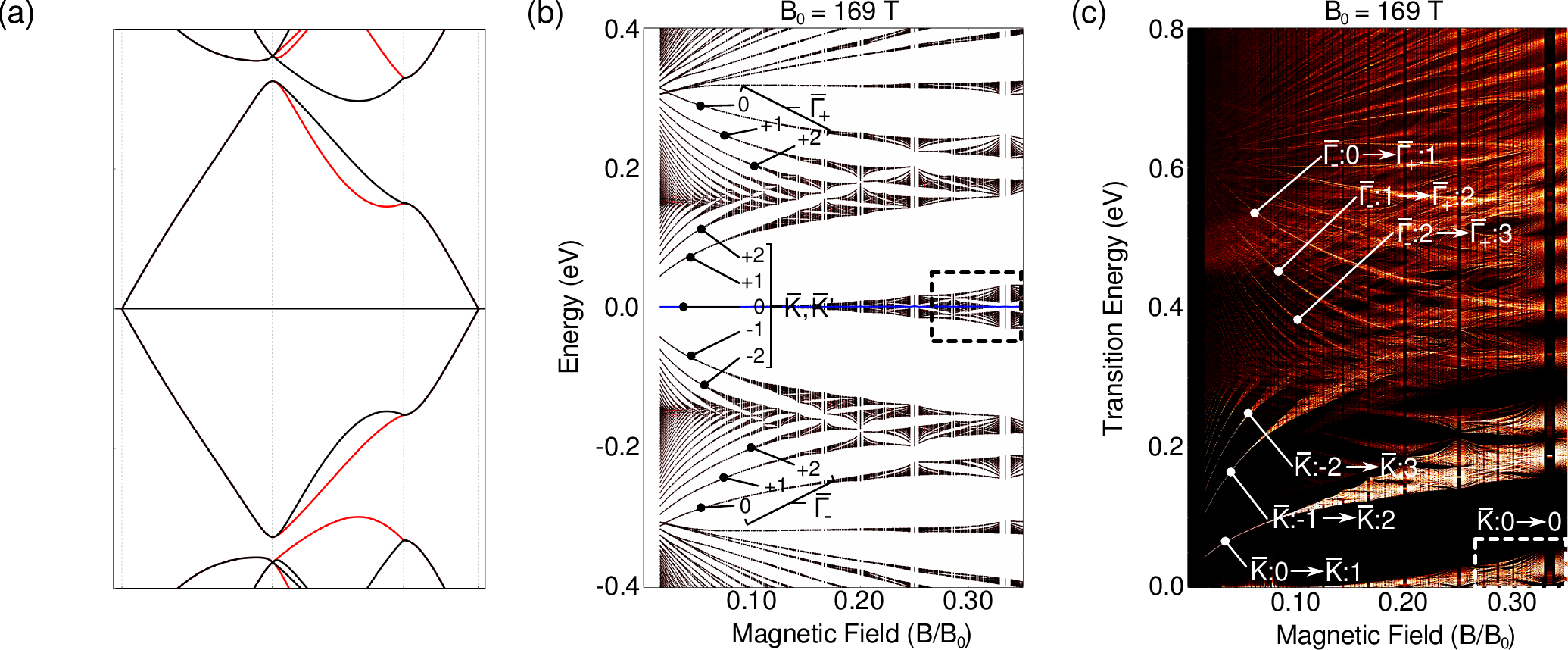}
\caption{(Color online) The electronic and magneto-optical properties of TBG with a twist angle of $2.65\,\mathrm{deg.}$ (a) The band structure at $B=0\,\mathrm{T}$. The black and red lines correspond to the bands originating from the $K$ and $K'$ valleys, respectively. (b) Energy spectrum as a function of magnetic field, $B$. The black and red lines correspond to the bands originating from the $K$ and $K'$ valleys, respectively. The blue line indicates the charge neutral point. The labels mark various Landau levels with their index, $n$ and the charge pocket from which they originate indicated. (c) The magneto-optical spectrum for left circularly polarized light ($\sigma_{+}$). The labels mark various transitions with their initial and final index, $n$, and the charge pocket from which they originate indicated. The collection of low energy transitions highlighted by the white dashed box originate from transitions between the sub-bands of the $n=0$ Landau level, which is marked in the energy spectrum in (b) by the black dashed box.}
\label{fig3}
\end{figure}

TBG displays many of the same features as MLG. The features of this spectrum were discussed in detail in Ref. \cite{tbgmo} - where a tight-binding approach was used to model the material - and, therefore, we will only briefly review the key aspect here. In the absence of a magnetic field the TBG band structure, shown in Fig.~\ref{fig3}(a), displays linear bands emanating from Dirac cones located at the corners of the moir\'{e} Brillouin zone. The band structure in the presence of a magnetic field is shown in Fig.~\ref{fig3}(b). At weak magnetic fields the continuous band structure splits up into the same series of discrete Landau levels. However, owing to the reduction in Brillouin zone size one can observe both the electron-like Landau levels (or hole-like Landau levels at -ve energies) associated with the $0\,\mathrm{eV}$ charge pockets at $\bar{K}$ and $\bar{K'}$ and the hole-like Landau levels (or electron-like Landau Levels at -ve energies) associated with the $0.25\,\mathrm{eV}$ charge pocket at $\bar{\Gamma}$. At higher magnetic fields, the bands evolve to display a more complicated fractal structure known as the Hofstadter spectrum. The transition between the semi-classical Landau level structure and the Hofstadter spectrum occurs when the magnetic length, $l_{B}$, becomes smaller than the moir\'{e} lattice constant, $L_{M}$, which occurs at about $B/B_{0} = p/q = \sqrt{3}/4\pi \approx 0.14$. In this regime the uncertainty of the momentum of the electron becomes comparable to the size of the Brillouin zone and hence quantum interference effects dominate. 

Figure \ref{fig3}(c) shows the magneto-optical spectrum for left circularly polarized light, which concurs with previous work \cite{tbgmo} (note that the magneto-optical spectrum for right circularly polarized light is almost identical to that of left circularly polarized light). At low magnetic fields and low transition energies the optical-conductivity obeys the same selection rules as MLG. One sees clear transition peaks between Landau levels of the $\bar{K}$ and $\bar{K'}$ charge pockets, which increase in energy as the magnetic field is increased. At higher energies one sees transition peaks between Landau levels of the $\bar{\Gamma}$ charge pocket. These decrease in energy as magnetic field increases. Transitions between the  $\bar{K}$, $\bar{K'}$ and $\bar{\Gamma}$ charge pockets are negligible because the momentum of the photon is much smaller than the momentum of the electrons and hence the $k$ of the electron is only negligibly changed in the excitation process. At higher energies the magneto-optical conductivity becomes more complicated as more energy transitions become possible.

At higher magnetic fields the individual Landau levels split into $2p$ subbands, where $p$ is given by $B/B_{0} = p/q$ and the $2$ refers to the number of layers. This fractal band structure leads to the appearance of a fractal structure for the magneto-optical conductivity and is a directly observable consequence of the appearance of the Hofstadter spectrum. Each subband is identified with a different, additional angular momentum component and hence the angular momentum of the subbands are now given by $L_{n} = \hbar(|n|+|m|)$, where $m$ is the index for the second-generation Landau levels. As a result, transitions within Landau level $n$ are now allowed because the difference in angular momentum of the subbands allows the material to absorb the angular momentum of the photon. This leads to the very low energy transitions observable in the bottom left of Fig.~\ref{fig3}(c) \cite{tbgmo}. Reference \cite{tbgmo} also highlighted the appearance of transitions between $|n| \rightarrow |n| + 3m \mp 1$ owing to trigonal warping effects. As each of the graphene layers is modelled by a low energy linear Hamiltonian these higher order effects are absent in the effective model presented here but are easily included by considering the low energy expansion of next-nearest neighbour couplings.

The TBG lattice obeys both $C_{2x}$ and $C_{2y}$ symmetry. The former ensures that the Landau level energies from a magnetic field $B$ are the same as those from a magnetic field $-B$. The latter ensures that the Landau level energies in one valley under a magnetic field $B$ are the same as those from a magnetic field $-B$ in the other valley. Together these symmetries ensure that one has valley degenerate Landau levels. In addition, the TBG Hamiltonian obeys the symmetry
\begin{equation}
\Sigma^{-1}H_{TBG}^{(\xi)}\Sigma = -H_{TBG}^{(-\xi)},
\label{sym}
\end{equation}
where
\begin{equation}
\Sigma = \left(\begin{array}{cc}
0 & \sigma_{x}\\
-\sigma_{x} & 0
\end{array}\right),
\end{equation}
and, hence, the band structure displays particle-hole symmetry (although in real materials this symmetry is broken by higher order effects and, hence, may not be precisely satisfied). Particle-hole symmetry mean that circular dichroism weak in TBG because the left circularly polarized light transition between $-n \rightarrow n+1$ has a similar transition energy to the right circularly polarized light transition  between $-(n+1) \rightarrow n$. Breaking of exact particle-hole symmetry will lead to larger circular dichroism.

\subsection{Twisted Double Bilayer Graphene}

\begin{figure}[ht]
\centering
\includegraphics[width=0.75\columnwidth]{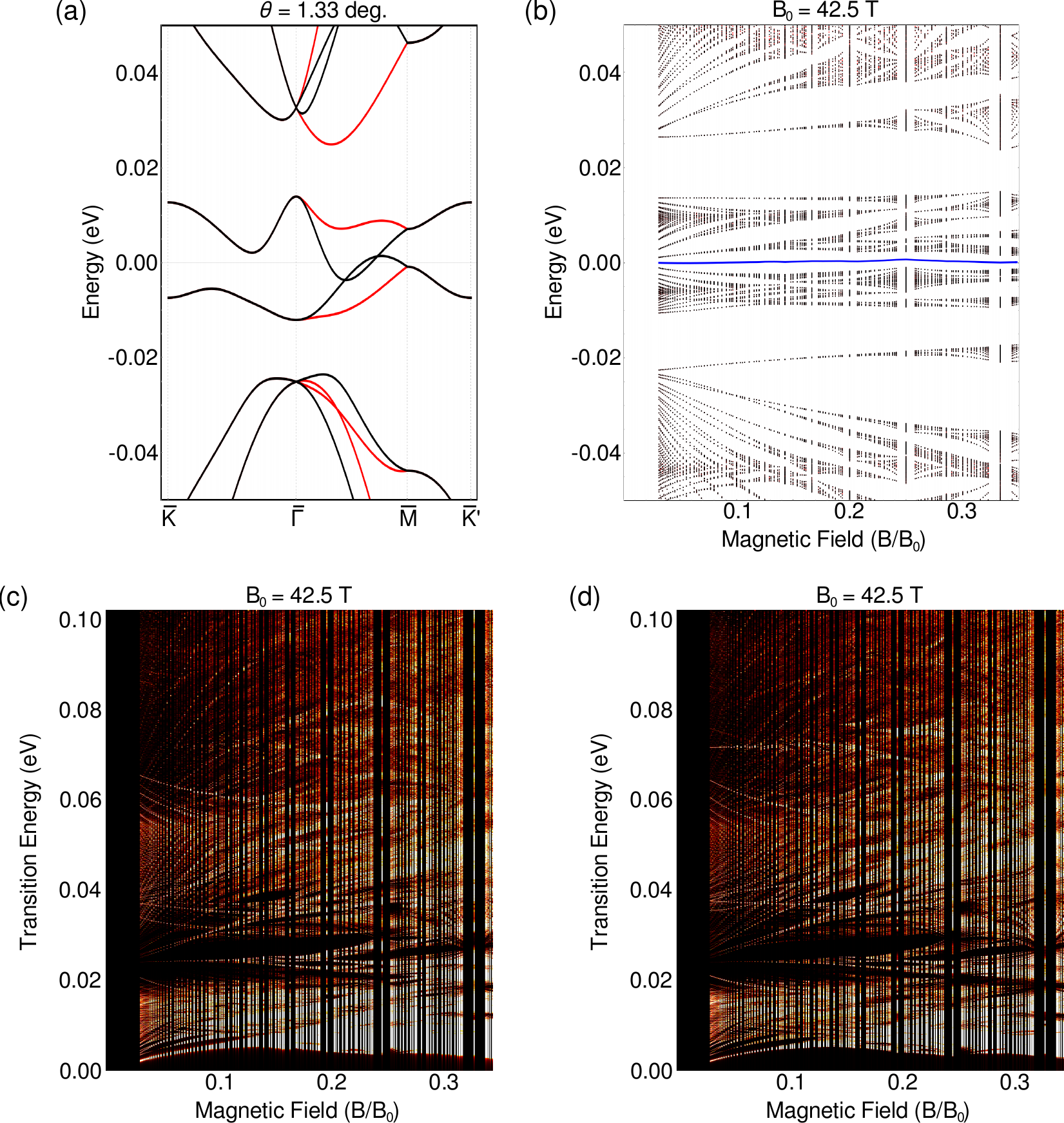}
\caption{(Color online) The electronic and magneto-optical properties of AB-AB TDBG (a) The band structure at $B=0\,\mathrm{T}$. The black and red lines correspond to the bands originating from the $K$ and $K'$ valleys, respectively. (b) Energy spectrum as a function of magnetic field, $B$. The black and red lines correspond to the bands originating from the $K$ and $K'$ valleys, respectively. The blue line indicates the charge neutral point. (c) and (d) The magneto-optical spectrum for left ($\sigma_{+}$) and right ($\sigma_{-}$) circularly polarized light, respectively.}
\label{fig4}
\end{figure}

TDBG has two different stackings, AB-AB and AB-BA, the latter of which is related to the former by a 180 degree in-plane rotation of one of the upper layer (see Fig.~\ref{fig2}). Despite having a similar atomic structure and similar band structure at $B=0\,\mathrm{T}$ their energy spectra under the influence of a magentic field (and, hence, their magneto-optical conductivites) differ dramatically. The reason for the difference is related to the symmetries of the underlying lattice. AB-AB stacked TDBG obeys $C_{2x}$ symmetry, which ensures that the energy spectrum under positive and negative B fields are identical. Time reversal symmetry ensures that the energy spectrum of one valley under a positive B field is the same as that of the opposing valley under a negative B field. The combination of these two symmetries means that the energy spectrum of AB-AB stacked TDBG is valley degenerate. In contrast, AB-BA stacked TDBG obeys $C_{2y}$ symmetry, which, like time reversal symmetry, ensures that the energy spectrum of one valley under a positive B field is the same as that of the opposing valley under a negative B field. Thus, valley degeneracy is not symmetry protected and one sees a dramatic difference between the energy spectrum of the two valleys \cite{tdbghof}. As a result the magneto-optical spectrum of AB-AB and AB-BA TDBG are also different.

Bilayer graphene also introduces a number of new effects that are not present in monolayer graphene. The parameter, $v_{4}$, in the Hamiltonian of both forms of TDBG leads to electron-hole asymmetry. The appearance of this term in the Hamiltonian induces a self coupling in each Landau level resulting in a level dependent energy shift. Thus, the energy of the transitions between $-n \rightarrow n+1$ driven by left circularly polarized light as and the transitions between $-(n+1) \rightarrow n$ driven by right circularly polarized light transition have different energies ensuring that circular dichroism is present. The parameter also allows for a new series of transitions within a single Landau level and, hence, transitions with the selection rule $|n| \rightarrow |n|$ are now allowed, even outside the Hofstadter spectrum regime. The parameter, $v_{3}$, leads to trigonal warping. This also directly affects the selection rules as this term in the Hamiltonain hybridizes the $|n|$ and $|n|+3$ Landau levels \cite{trigwarp}. As a result, an extra set of transitions between $|n| \rightarrow |n| + 3m \mp 1$, are now allowed.

First consider AB-AB stacked TDBG. Unlike TBG, at low magnetic fields TDBG does not display a clear set of regularly spaced Landau levels. This is because the $B=0\,\mathrm{T}$ band structure does not display the simple linear or parabolic structure one finds in other materials. This can be seen in Fig.~\ref{fig4}(a). Figure \ref{fig4}(b) shows the band structure for AB-AB TDBG in the presence of a magnetic field. The central region of Landau levels between $-10.5\,\mathrm{meV} < E < 13.6\,\mathrm{meV}$ originate from the charge pockets around both $\bar{K}$, $\bar{K'}$ and $\bar{\Gamma}$ whilst the Landau levels at $E < -22.6\,\mathrm{meV}$ and $E > 26.4\,\mathrm{meV}$ originate from charge pockets located near, but not at, $\bar{\Gamma}$. Figures \ref{fig4}(c) and (d) show the magneto-optical spectrum for left and right circularly polarized light, respectively. As the small momentum of the photon cannot drive transitions between different values of $k$, the high energy transitions are exclusively caused by transitions near $\bar{\Gamma}$ whereas the lower energy transitions can occur at any of the high symmetry points $\bar{K}$, $\bar{K'}$ and $\bar{\Gamma}$. The main feature of the low magnetic field regime is a comparative absence of optical transitions in the region between $20\,\mathrm{meV}$ and $40\,\mathrm{meV}$. The large collection transitions at lower energies stem from transitions within the central region of densely packed Landau level. The higher energy transitions come exclusively from transitions between the Landau levels at $E < -22.6\,\mathrm{meV}$ and those in the central region or those in the central region and the Landau levels at $E > 26.4\,\mathrm{meV}$. Both the left circularly driven $|n| \rightarrow |n|+1$ transitions and right circularly driven $|n| \rightarrow |n|-1$ transitions are of similar but non-identical in energy. This leads to the small amount of circular dichroism the material displays. At higher magnetic fields the magneto-optical conductivity shows a complex fractal structure reminiscent of the Hofstadter spectrum. As the magnetic field is increased the gaps that are present above and below the central region of Landau levels narrows and hence the energy range of potential transition close to this gap increases. Thus, one sees the appearance of transitions in the region $20\,\mathrm{meV}$ and $40\,\mathrm{meV}$ which was previously devoid of strong transitions at low magnetic fields.

\begin{figure}[h]
\centering
\includegraphics[width=0.75\columnwidth]{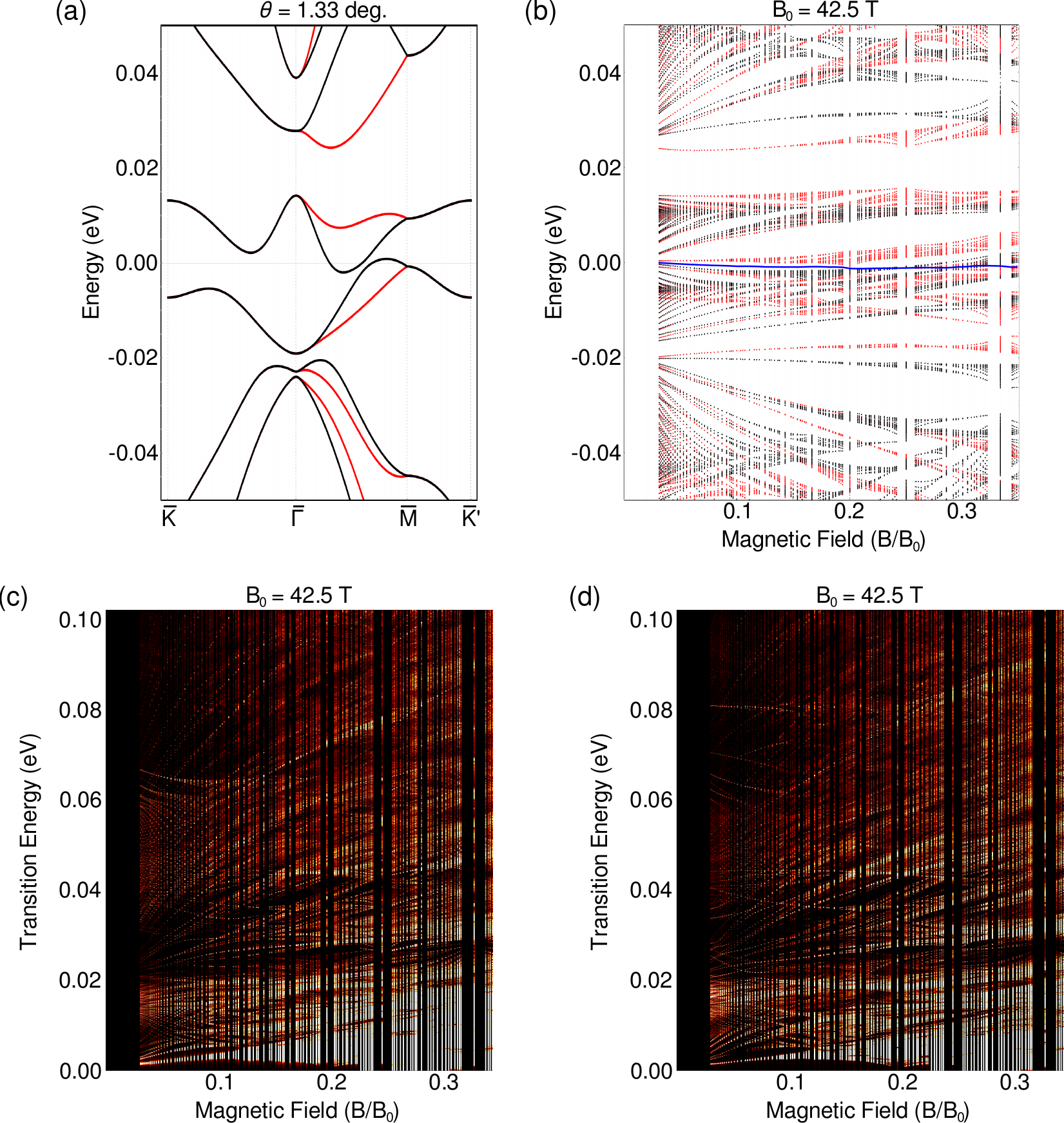}
\caption{(Color online) The electronic and magneto-optical properties of AB-BA TDBG (a) The band structure at $B=0\,\mathrm{T}$. The black and red lines correspond to the bands originating from the $K$ and $K'$ valleys, respectively. (b) Energy spectrum as a function of magnetic field, $B$. The black and red lines correspond to the bands originating from the $K$ and $K'$ valleys, respectively. The blue line indicates the charge neutral point. (c) and (d) The magneto-optical spectrum for left ($\sigma_{+}$) and right ($\sigma_{-}$) circularly polarized light, respectively.}
\label{fig5}
\end{figure}
Next consider AB-BA stacked TDBG. Again, unlike TBG, and for the same reasons [see Fig.~\ref{fig5}(a)], at low magnetic fields TDBG does not display a clear set of regularly spaced Landau levels. Figure \ref{fig5}(b) shows the band structure for AB-BA TDBG in the presence of a magnetic field. Note that, here, there is a dramatic difference between the bands from the $\bar{K}$ and $\bar{K'}$ valleys. As before, the central region of Landau levels between $-16.5\,\mathrm{meV} < E < 14.1\,\mathrm{meV}$ originate from the charge pockets around both $\bar{K}$, $\bar{K'}$ and $\bar{\Gamma}$ whilst the Landau levels at $E < -19.8\,\mathrm{meV}$ and $E > 24.0\,\mathrm{meV}$ originate from charge pockets located near, but not at, $\bar{\Gamma}$. Figures \ref{fig4}(c) and (d) show the magneto-optical spectrum for left circularly polarized light and right circularly polarized light, respectively. The features seen are similar to that of AB-AB stacked TDBG but with the gap region between $20\,\mathrm{meV}$ and $40\,\mathrm{meV}$ not so distinct. This is owing to a central region of Landau levels that is broader in energy than in AB-AB TDBG. Again, the large collection transitions at lower energies stem from transitions within the central region of densely packed Landau level and the higher energy transitions come from transitions either from $E < -19.8\,\mathrm{meV}$ to the central region or from the central regions to bands above $E > 24.0\,\mathrm{meV}$. At higher magnetic fields the magneto-optical conductivity, once again, shows a complex fractal structure reminiscent of the Hofstadter spectrum.

\subsection{Monolayer Graphene - Hexagonal Boron Nitride}

\begin{figure}[ht]
\centering
\includegraphics[width=0.75\columnwidth]{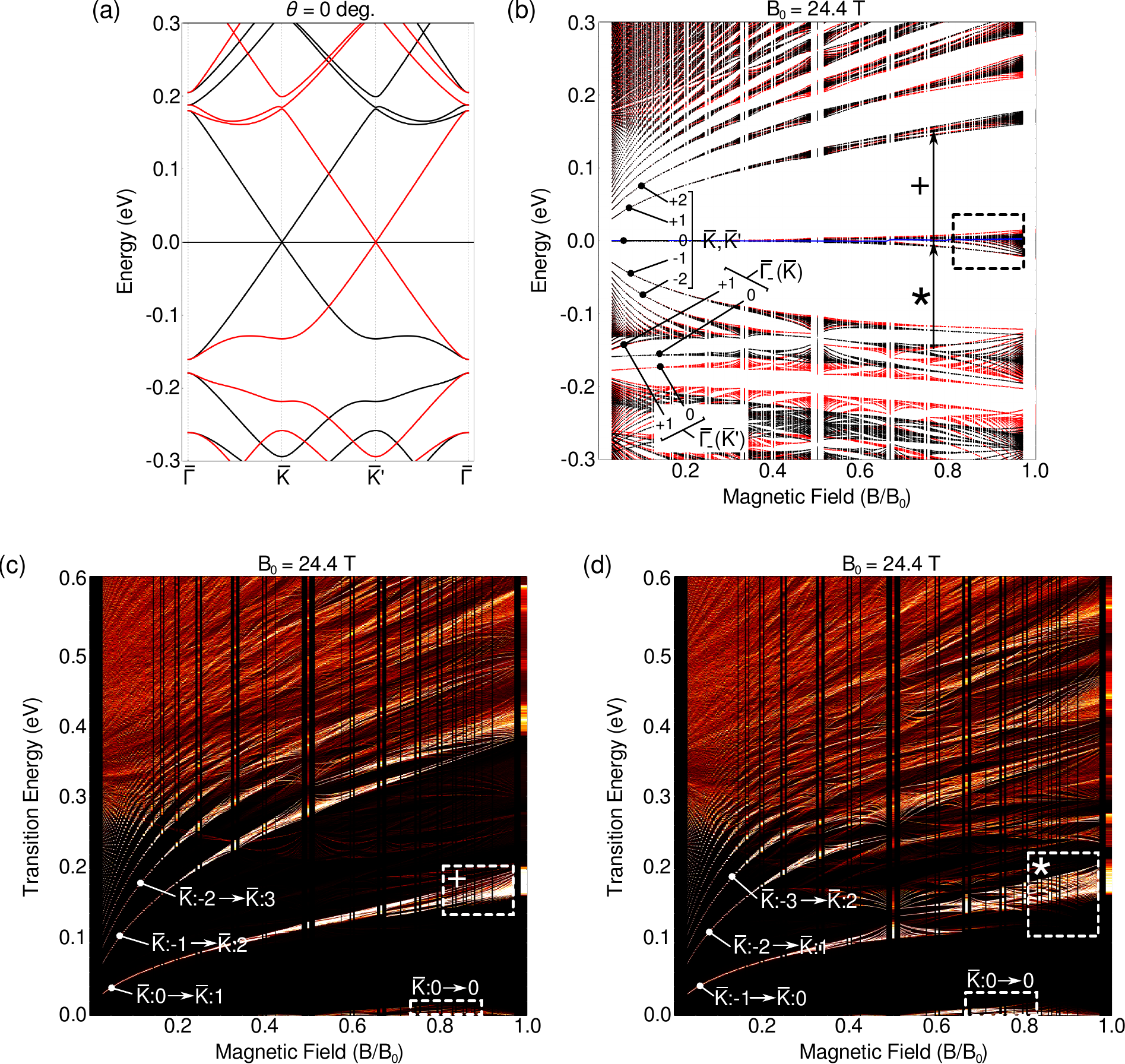}
\caption{(Color online) The electronic and magneto-optical properties of MLG-hBN (a) The band structure at $B=0\,\mathrm{T}$. The black and red lines correspond to the bands originating from the $K$ and $K'$ valleys, respectively. (b) Energy spectrum as a function of magnetic field, $B$. The black and red lines correspond to the bands originating from the $K$ and $K'$ valleys, respectively. The blue line indicates the charge neutral point. The labels mark various Landau levels with their index, $n$, and the charge pocket from which they originate indicated. The transitions marked $+$ and $\ast$ are the transitions that most significantly contribute to optical dichroism. (c) and (d) The magneto-optical spectrum for left ($\sigma_{+}$) and right ($\sigma_{-}$) circularly polarized light, respectively. The labels mark various transitions with their initial and final index, $n$ and the charge pocket from which they originate indicated. The collection of low energy transitions highlighted by the white dashed box originate from transitions between the sub-bands of the $n=0$ Landau level, which is marked in the energy spectrum in by the black dashed box. The energy regions marked $\ast$ and $+$ are the transitions that most significantly contribute to optical dichroism and correspond to the transitions marked in (b).}
\label{fig6}
\end{figure}

MLG-hBN is a heterogeneous material, which means that, unlike TBG and TDBG, the lattice obeys neither $C_{2x}$ nor $C_{2y}$ symmetry. Furthermore, the potential from the hBN layer does not obey the symmetry in Eq. \eqref{sym}. Hence, neither valley degeneracy nor particle-hole symmetry are enforced. The lack of both valley degeneracy and particle-hole symmetry can be clearly seen in Figs.~\ref{fig6}(a) and (b), which show, respectively, the zero-field band structure and the band structure under the influence of a magnetic field. In this case one would expect to observe both significant circular and valley dichroism. Figures \ref{fig6}(c) and (d) show the magneto-optical conductivity for left and right circularly polarized light and, as expected, they differ significantly.

The magento-optical spectrum, again, displays strong peaks at energies which are resonant with transitions between Landau levels. The effective theory models the interaction with the hBN layer using the potential in Eq. \eqref{VhBN} which is derived from the interlayer interaction, Eq. \eqref{U}, and the dispersionless low energy Hamiltonian for the hBN layer, Eq. \eqref{hBN}. Thus, the potential in Eq. \eqref{VhBN} is not a function of the electron momentum and, hence, does not contribute to the magneto-optical conductivity directly - the potential only acts to change energies of the Landau levels themselves. Thus, the selection rules for this material are the same as for monolayer graphene - $|n| \rightarrow |n|+1$ for left circularly polarized light and $|n| \rightarrow |n|-1$ for right circularly polarized light. As the graphene layer is described by the linear low energy Hamiltonian the $|n| \rightarrow |n| + 3m \mp 1$ transitions from trigonal warping effects are absent. The similarities with monolayer graphene are particularly apparent at low energies and low magnetic fields where neither the difference between the particle and hole bands or the difference between the bands originating at the two valleys are particularly pronounced. This is because, in this regime, the band structure is dominated by the low energy bands from the graphene layer, rather than the bands from the hBN layer whose energy is significantly higher. Here, circular dichroism is small and the spectrum resembles that of monolayer graphene. Such phenomenology was observed in Ref. \cite{ghbnabs} with the minor deviations of the experimental data from the theoretical monolayer graphene bands probably a result of the effect of the hBN layer potential. Note that the strong optical phonon mode of hBN at $E\approx 0.17\,\mathrm{eV}$ observed in Ref. \cite{ghbnabs} is not included in the calculation and in reality one would expect to observe a similar discontinuity in both here and in BLG-hBN, which is discussed below. As the energy increases the difference between the particle and hole bands becomes more dramatic as the influence of the hBN layer increases. However, as the energy increases the number of bands becomes large and hence the difference is obscured by the shear number of possible transitions. 

At higher magnetic fields the circular dichroism becomes more discernable as the gaps between the Landau levels broaden in energy. The distinction becomes pronounced at $B/B_{0} = p/q \approx 0.4$ when the fractal spectrum of the valence bands becomes fully developed. The most dramatic difference occurs for the left circularly polarized $0 \rightarrow 1$ transition [marked by $+$ in Fig.~\ref{fig6}(b) and (c)] compared to the right circularly polarized $-1 \rightarrow 0$ transition [marked by $\ast$ in Fig.~\ref{fig6}(b) and (d)]. In the former one sees a narrower collection of transitions compared to the broader collection for the latter, the difference stemming from the particle-hole asymmetry of the band structure. 

\subsection{Bilayer Graphene - Hexagonal Boron Nitride}

\begin{figure}[h]
\centering
\includegraphics[width=0.75\columnwidth]{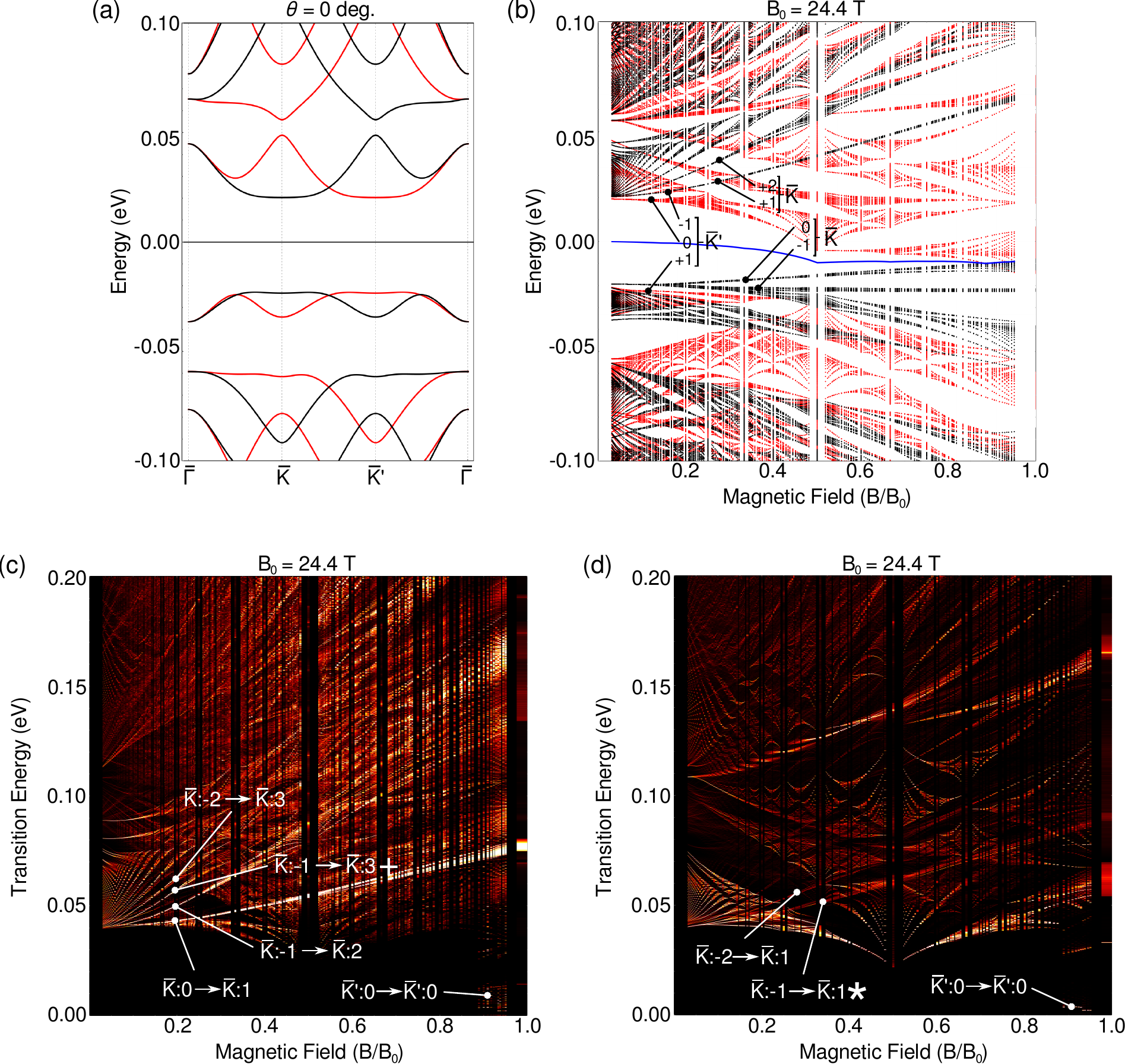}
\caption{(Color online) The electronic and magneto-optical properties of BLG-hBN (a) The band structure at $B=0\,\mathrm{T}$. The black and red lines correspond to the bands originating from the $K$ and $K'$ valleys, respectively. (b) Energy spectrum as a function of magnetic field, $B$. The black and red lines correspond to the bands originating from the $K$ and $K'$ valleys, respectively. The blue line indicates the charge neutral point. The labels mark various Landau levels with their index, $n$, and the charge pocket from which they originate indicated. (c) and (d) The magneto-optical spectrum for left ($\sigma_{+}$) and right ($\sigma_{-}$) circularly polarized light, respectively. The labels mark various transitions with their initial and final index, $n$, and the charge pocket from which they originate indicated. The transitions marked $+$ and $\ast$ in (c) and (d) show a transition owing to trigonal warping effects and the appearance of subbands respectively.}
\label{fig7}
\end{figure}
BLG-hBN, like MLG-hBN is a heterogeneous material which is subject to the potential from the hBN layer. This means that, again, the lattice obeys neither $C_{2x}$ nor $C_{2y}$ symmetry nor the symmetry in Eq. \eqref{sym} and, hence, neither valley degeneracy nor particle-hole symmetry are enforced, as can be seen from Fig.~\ref{fig7}(a) and (b). However, the presence of the extra graphene layer means that both the particle-hole and valley asymmetries are far more pronounced. This is because the hBN potential only affects one of the two layers in the bilayer and hence inversion symmetry of the bilayer is more strongly broken \cite{ghbnhof}. The result is a complicated energy spectrum when a magnetic field is applied with a striking differences between the contribution from the two valleys. At low magnetic fields the $\bar{K}$ valley displays Landau level like band structure whereas the $\bar{K}'$ valley already displays a fractal-like bands. The reason for this is that the wavefunction of the zero energy Landau level is localized on upper layer of the graphene bilayer in the $\bar{K}$ valley and on the lower layer of the graphene bilayer in the $\bar{K}'$ valley. Hence, the bands in the $\bar{K}'$ valley are much more strongly affected by the presence of the hBN layer adjacent to the lower layer and, thus, the hofstadter spectrum develops more rapidly in this valley.

The presence of the bilayer also introduces many features that were previously found in both AB-AB and AB-BA TDBG. The interlayer interaction within the bilayer introduces both trigonal warping and particle-hole symmetry breaking via the $v_{3}$ and $v_{4}$ terms of Eq. \eqref{BLGint}, respectively, the former of which leads to the $|n| \rightarrow |n|$ and $|n| \rightarrow |n| + 3m \mp 1$ transitions previously discussed in the context of TDBG \cite{trigwarp}. Furthermore, unlike MLG-hBN, which is gapless at $B=0\,\mathrm{T}$, BLG-hBN exhibits a large gap of around $\approx 40\,\mathrm{meV}$. This is caused by the proximity of the hBN layer, which induces a potential in the adjoining layer resulting in a potential asymmetry across the two BLG layers. This, clearly, results in an absence of low energy transitions across most of the magneto-optical spectrum until the increasing magnetic field raises the valence band energy in the positive valley (and lowers the conduction band energy in the negative valley) such that the bands rise above (drops below) the charge neutral point, at which point low energy transitions between the subbands of the Hofstadter spectrum become feasible. This occurs at $p/q \approx 0.9$. As the energy increases, one observes the appearance of transitions peaks once the energy is enough to drive transitions across the gap. At low magnetic fields one sees the distinct peaks associated with transitions between individual landau levels. However, in this case these peaks do not display the usual $\propto \sqrt{B}$ dependence as the bands around the $\bar{K}$ and $\bar{K}'$ are almost flat as opposed to linear as in MLG (or MLG-hBN). At higher magnetic fields the Landau levels obtain a linear characteristic reminiscent of non-relativistic Landau levels from parabolic bands, before developing into a the more complex fractal structure of the Hofstadter spectrum.

Figures \ref{fig7}(c) and (d) show the magneto-optical spectrum for left and right circularly polarized light, respectively. Both show a complicated series of transitions that are a combination of the Landau level like bands originating from the $\bar{K}$ valley and the fractal like bands originating from the $\bar{K}'$ valley. Thus, unlike the case of MLG-hBN, where the valley dichroism was small owing to the small separation of the transition peaks, BLG-hBN offers the opportunity for optically induced valley polarization since there is a significant difference between the spectra of the two valleys and there are large regions in both spectra where the dominant transitions originate from a single valley. Note, again, that the discontinuity in the magneto-optical spectrum caused by the strong optical phonon mode of hBN at $E\approx 0.17\,\mathrm{eV}$, which was observed in Ref. \cite{ghbnabs}, is not included in the calculation but one would expect to observe it here as well as in MLG-hBN, which was discussed above.

Figures \ref{fig7}(c) and (d) also highlight a number of the strong transitions. At low magnetic fields, left circularly polarized light displays a number of strong, narrow transitions that stem from $\bar{K}$ valley with the the $0 \rightarrow 1$ transition in the being particularly prominent. One also sees a strong $-1 \rightarrow 3$ transition, [marked with a $+$ in Fig.~\ref{fig7}(c)] that is allowed via the $v_{3}$ trigonal warping term in the Hamiltonian. The transition obeys the $|n| \rightarrow |n| + 3m - 1$ selection rule with $m=1$. At higher magnetic fields left circularly polarized light still displays prominent strong transitions, though broadened by the development of the fractal hofstadter spectrum. One also sees the appearance of low energy $0 \rightarrow 0$ transition which appear owing to the closing of the gap at high magnetic fields along with the appearance of hofstadter subbands. Conversely, right circularly polarized light displays a large number of weaker transitions that stem from the $\bar{K}'$ valley and a few weak transitions that stem from the $\bar{K}$ valley. A notable transition here is the $-1 \rightarrow 1$ transition [marked with a $\ast$ in Fig.~\ref{fig7}(c)] that is allowed because of the appearance of the Hofstadter subbands.

\subsection{Circular Dichroism, Faraday Rotations and Ellipticity}

\begin{figure}[h]
\centering
\includegraphics[width=0.75\columnwidth]{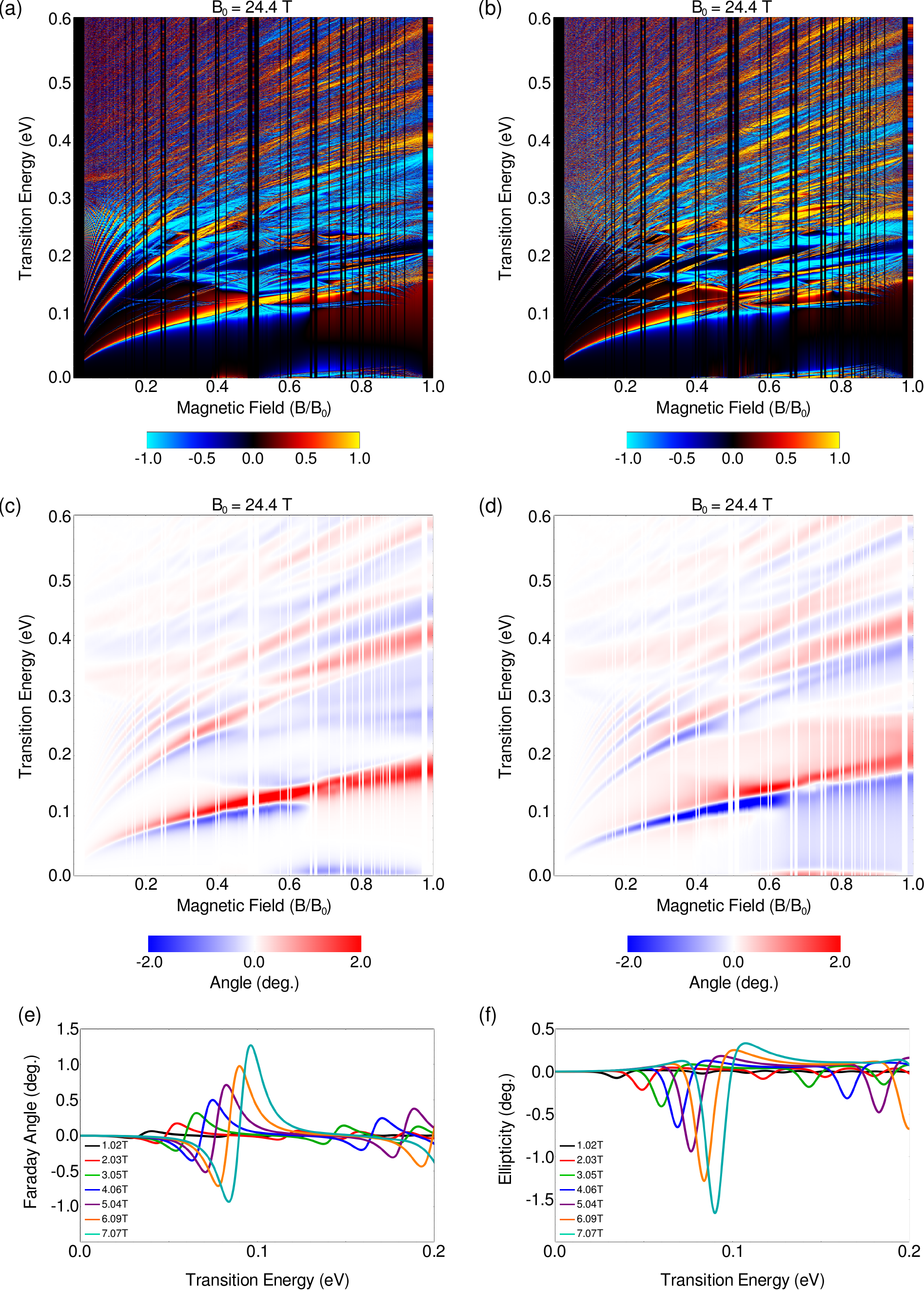}
\caption{(Color online) (a) Circular and (b) valley dichroism for MLG-hBN. (c) Faraday rotation angle and (d) Ellipticity. Low energy (e) Faraday rotation and (f) ellipticity for a variety of magnetic field values. Note that for (c)-(f) the phenomenological broadening parameter is taken to be $10~\mathrm{meV}$.}
\label{fig8}
\end{figure}
Owing to the strong breaking of inversion symmetry by the hBN layer both MLG-hBN and BLG-hBN display distinct spectra when illuminated by left or right circularly polarized light. Figure \ref{fig8}(a) shows the circular dichroism ($D_{c}$) for MLG-hBN as defined by Eq. \eqref{cd}. At low energies and low magnetic fields one sees a slight circular dichroism owing to the slight difference in energy of the particle-like and hole-like bands. Significant circular dichroism over a large energy region/magnetic field range is only present at higher energies and higher magnetic fields when the differences between the particle and hole bands becomes pronounced.

The dichroism characteristics of the material can be probed by measuring the Faraday rotation and ellipticity of transmitted plane polarized light, whose magnitude is related to the difference in the real and imaginary parts of the right and left circularly polarized magneto-optical conductivity via Eqs. \eqref{theta} and \eqref{delta}, respectively. Figure \ref{fig8}(c) show the Faraday rotation angle and Fig.~\ref{fig8}(d) show the ellipticity as a function of magnetic field and transition energy. Here we have used a phenomenological broadening of $\eta = 10\,\mathrm{meV}$ (rather than the value of $\eta = 0.01\,\mathrm{meV}$ used in to compute the magneto-optical conductivity) to match the observed disorder broadening in experiments on similar systems \cite{farad4}. This broadening washes out the details of the Hofstadter spectrum but leaves the overall shape intact. Figure \ref{fig8}(e) and (f) show the Faraday rotation angle ellipticity for low energies and a range of magnetic fields. Near the resonant transitions one observes large Faraday rotations and ellipticity on the order of a degree. 

Finally it is worth noting that, owing to the difference in the band structure for the $K$ and $K'$ valleys, valley dichroism will be present in this material. Figure \ref{fig8}(b) shows the valley dichroism ($D_{v}$) for MLG-hBN as defined in Eq. \eqref{vd}. Owing to minor differences in the band energies at the $K$ and $K'$ valleys one sees slight dichroism effects near the energies of the major transitions. However, the significant differences in the bands from the two valleys occurs at energies deep in the valence band side of the spectrum and mainly originates from the charge pockets at the $\bar{\Gamma}$-point. Thus, the transitions energies needed to observe the difference are large and, hence, the effect is somewhat obscured by the numerous high energy transitions between Landau levels originating from the $\bar{K}$-points. At higher magnetic fields one sees larger regions where a valley polarized response is possible although a large energy is still required to access these regions.

\begin{figure}[h]
\centering
\includegraphics[width=0.75\columnwidth]{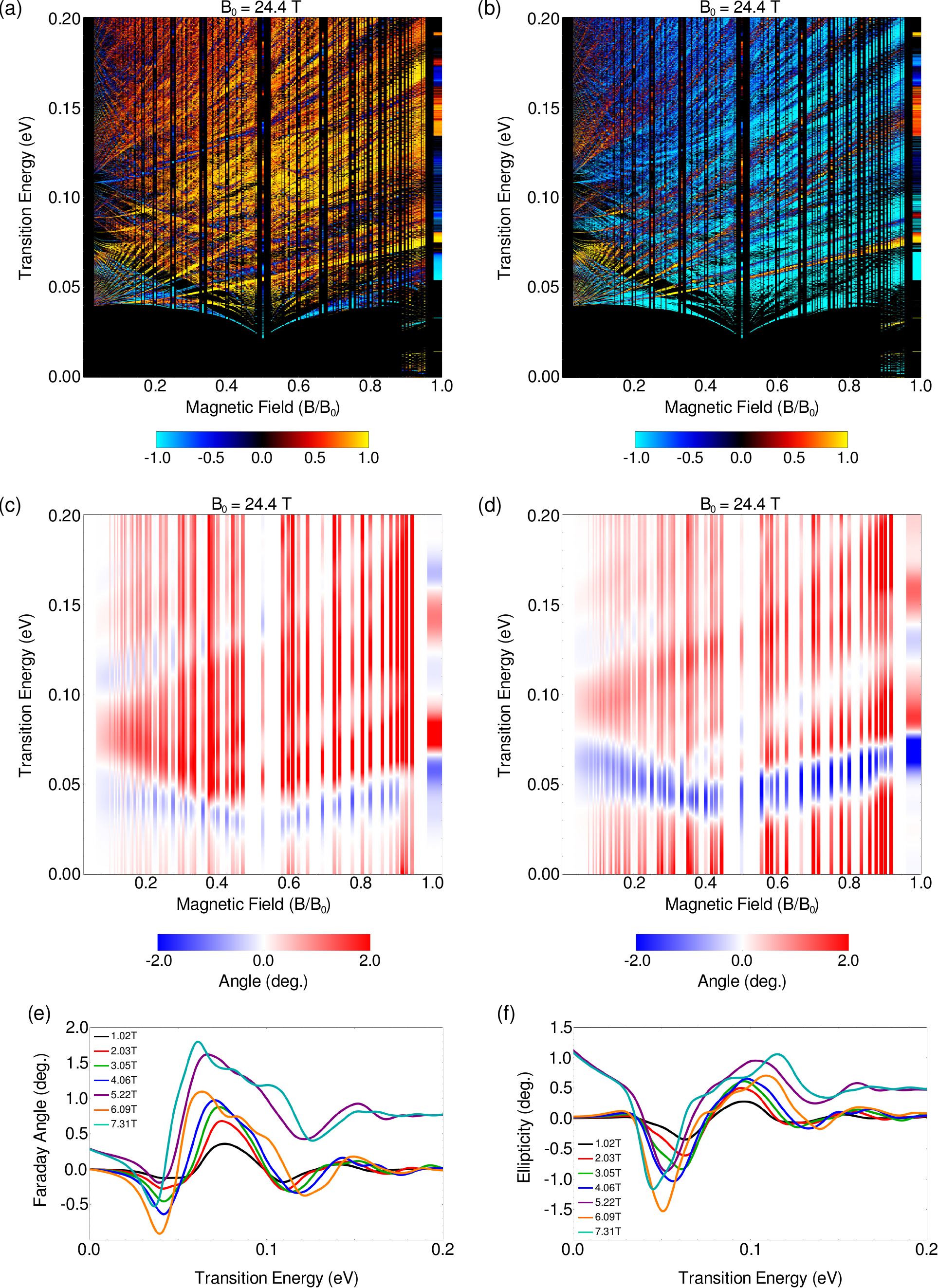}
\caption{(Color online) (a) Circular and (b) valley dichroism for BLG-hBN. (c) Faraday rotation angle and (d) Ellipticity. Low energy (e) Faraday rotation and (f) ellipticity for a variety of magnetic field values. Note that for (c)-(f) the phenomenological broadening parameter is taken to be $10\,\mathrm{meV}$.}
\label{fig9}
\end{figure}
Like its monolayer counterpart, BLG-hBN also displays inversion symmetry breaking and, hence, strong circular dichroism and likewise, displays a similar Faraday rotation and ellipticity when illuminated by plane polarized light. Again, Fig.~\ref{fig9}(a) shows the circular dichroism ($D_{c}$) for MLG-hBN as defined by Eq. \eqref{cd}. One can see for a large region where the interaction with right left circularly polarized light is strongly favoured. Figure \ref{fig9}(c) show the Faraday rotation angle and Fig.~\ref{fig9}(d) show the ellipticity as a function of magnetic field and transition energy with a phenomenological broadening of $\eta = 10\,\mathrm{meV}$\cite{farad4}, which, again, obscures the details of the Hofstadter spectrum. Figure \ref{fig9}(e) and (f) show the Faraday rotation angle ellipticity for low energies and a range of magnetic fields. The magnitudes of these effects is about the same order of magnitude as in MLG-hBN - about a degree.

Finally, Fig.~\ref{fig9}(b) shows the valley dichroism as defined by Eq. \eqref{vd}. The valley dichroism is significantly greater than that of MLG-hBN and is addressable over a wider range of energies and magnetic fields. Thus, BLG-hBN may offer a better platform for valleytronic applications compared to it's monolayer counterpart. 

\section{Summary}

Here we have studied the magneto-optical properties of five van der Waals heterostructures. In the low magnetic field regime all structures show transitions with selection rules $|n| \rightarrow |n| \pm 1$ for left and right circularly polarized light. Structures with bilayer graphene also show display $|n| \rightarrow |n| \pm 3m \mp 1$, $|n| \rightarrow |n|$ which is allowed via trigonal warping and electron-hole asymmetry effects, respectively. As TBG does not break inversion symmetry, it only displays weak circular or valley dichroism. In contrast ABAB-TDBG, ABBA-TDBG, MLG-hBN and BLG-hBN all break inversion symmetry and hence display significant circular dichroism and all but ABAB-TDBG also showing valley dichroism. Differences in the real and imaginary parts of the of the magneto-optical conductivity for left and right circularly polarized light lead to a Faraday rotation and ellipticity in incident plane polarized light. MLG-hBN and BLG-hBN both exhibit Faraday rotations and ellipticities of about a degree.

\section{Acknowledgements}

J.A.C was supported by the National Science Foundation of China Research Grant No. 12050410228.
P.M. acknowledges the support by
National Science Foundation of China (Grant No. \textcolor{red}{XXX}) and
Science and Technology Commission of
Shanghai Municipality (Shanghai Natural Science Grants,
Grant No. 19ZR1436400).
J.A.C. and P.M. were supported by
the NYU-ECNU Institute of Physics at NYU Shanghai. This
research was carried out on the High Performance Computing
resources at NYU Shanghai.

\end{CJK*} 
\end{document}